\documentclass[authoryear,preprint,review,12pt]{elsarticle}

\usepackage{amssymb}
\journal{New Astronomy}

\begin{document}

\begin{frontmatter}


\title{Frequency and spectrum analysis of  $\gamma$ Doradus type {\it Kepler} target KIC\,6462033}
%\thanks{Based on observations made with the Russian-Turkish telescope operated  T\"UB\.ITAK National Observatory,  Sakl{\i}kent, Antalya and the Mercator Telescope, operated on the island of La Palma by the Flemish Community, at the Spanish Observatorio del Roque de los Muchachos of the Instituto de Astrofísica de Canarias.}


\author[l1]{C. Ulusoy \corref{cer}}
\author[l2]{I. Stateva}
\author[l2]{I. Kh. Iliev}
\author[l3]{B. Ula{\c s}}
\address[l1]{College of Graduate Studies, University of South Africa,
PO Box 392, Unisa, 0003, Pretoria, South Africa \corref{cer}}
\address[l2]{Institute of Astronomy with NAO, Bulgarian Academy of Sciences,
blvd.Tsarigradsko chaussee 72, Sofia 1784, Bulgaria}
\address[l3]{{I}zmir Turk College Planetarium, 8019/21 sok., No: 22,
\.{I}zmir, Turkey}
\cortext[cer]{Corresponding author. Tel.: +27792420624; Fax: +274294672 \\
E-mail address: tulusoc@unisa.ac.za}

%\cortext[kyv]{Visiting astronomer during the summer of 2011}
\begin{abstract}
We present results of an asteroseismic study on the $\gamma$ Dor type {\it Kepler} target KIC\,6462033. {\it Kepler} photometry is used to derive the frequency content and principal modes. High-dispersion ground-based spectroscopy is also carried out in order to determine the atmospheric  parameters and projected rotational velocity. From an analysis of the {\it Kepler} long cadence time series, we find that the light curve of KIC\,6462033 is dominated by three modes with frequencies  $f_{1}$=0.92527, $f_{2}$=2.03656 and  $f_{3}$=1.42972 d$^{-1}$ as well as we detect more than a few hundreds of combination terms. However, two other independent frequencies appear to have lower amplitudes in addition to these three dominant terms. No significant peaks are detected in the region $>$ 5 d$^{-1}$. We therefore confirm that KIC\,6462033 pulsates in the frequency range of $\gamma$ Dor type variables, and a future study will allow us to investigate modal behaviour in this star.

\end{abstract}


\end{frontmatter}

\section{Introduction}
The $\gamma$ Dor stars have been known as a class of pulsating late A and F type variables for more than two decades. After the variability of the prototype, $\gamma$ Doradus,  was discovered by \cite{cowa63} they were first named and classified as a new group of pulsating stars by \cite{bal94} and \cite{kaye99}, respectively. Their oscillations are characterized by high-order, low-degree and multiple non-radial $g$-modes with periods of 0.3 to 3 d \citep{kaye99, bal11}.  

Since they have both convective cores and convective envelopes driving is assumed to be operated by a convective flux-blocking mechanism at the base of their convective envelope where radiative damping occurs in the $g$-mode cavity \citep{guz00, dup04, dupret05, gri05}. Morever, the depth of the convective envelope is play an important role to drive $g$-mode pulsations in $\gamma$ Dor stars, and the driving mechanism becomes efficient when the position of convective envelope makes the thermal relaxation time comparable to $g$-mode periods \citep{dup04, gri05, bal11}. These stars are located in a region on/or near the main sequence (with masses of 1.5 to 1.8 $M_\odot$ \citep{aerts10}) that partially overlaps with the red edge of $\delta$ Sct instability strip in the Hertzprung-Russell diagram (HRD). In this case, they are expected to show both $p$- and $g$-mode hybrid pulsations in their frequency spectra. From the ground-based observations $\gamma$ Dor/ $\delta$ Sct hybrid pulsations were first found by \cite{hand02}. Following that, a few hybrid type pulsator candidates were reported by several authors \citep{hefe05, ro06, king07}. By using  {\it Kepler} data, \cite{gri10} have recently proposed a new observational classification scheme for the pulsators located in this overlapping region.  

In particular, new generation space missions such as {\it Kepler} \citep{bor10}, {\it CoRoT} \citep{bag06} and {\it MOST} \citep{wal03} allow us to detect further hybrid candidates with very low-amplitude pulsation modes which may challenge our understanding of their unique pulsation mechanism \citep{har10, uyt11, tkac13}.

The star KIC\,6462033 (TYC 3144-646-1, $V= 10.83$, $P=0.69686$ d) has been monitored by the {\it Kepler} satellite in short-cadence (SC, 1-min exposures, \cite{gil10}) and long cadence (LC, 29.4-min exposures, \cite{jen10}) modes, and was firstly classified as a $\gamma$\,Dor-type star by \cite{uyt11}. 

In this paper, we present results of a frequency analysis of the  {\it Kepler}  LC dataset to investigate pulsational frequencies as well as an analysis of ground-based spectra.  {\it Kepler} observations and data processing procedure and the frequency analysis are described in Section 2. Section 3 deals with the derivation of the fundamental stellar parameters from the ground-based spectroscopy. Conclusions are briefly discussed in Section 4. 

\begin{table}
\centering
\caption{Frequencies, amplitudes ($A$), phases ($\phi$)  and $S/N$ for the combined (LC) Q0-Q16 data set of KIC\,6462033. Uncertainities are presented in paranthesis.}
\label{tabfourier}
\begin{tabular}{lrrrr}
\hline
\hline
ID&$f$(d$^{-1}$)& $A$(mmag) & $\phi$ (radians)  & $S/N$ \\
\hline
$f_{1}$ & 0.92527(2) & 0.821(19) & 5.785(11) & 74 \\
$f_{2}$ & 1.42972(2) & 0.696(16) & 4.248(11) & 76 \\
$f_{3}$ & 2.03656(2) & 0.583(17) & 2.888(14) & 58 \\
$f_{4}$ & 1.02317(2) & 0.523(13) & 3.707(11) & 70 \\
$f_{5}$ & 3.43850(2) & 0.481(12) & 5.382(11) & 71 \\
$f_{6} \approx 2f_{4}$ 	& 2.04634(3) & 0.261(9)  & 4.232(17) & 48 \\
$f_{7} \approx 3f_{4}$ 	& 3.06955(9) & 0.028(3)  & 4.745(59) & 14 \\
$f_{8} \approx 2f_{1}$ 	& 1.85030(15) & 0.023(5) & 4.570(103) & 8 \\
$f_{9} \approx 2f_{5}$ 	& 6.87700(13) & 0.015(3) & 3.849(87) & 9 \\
$f_{10} \approx 4f_{4}$ 	& 4.09270(12) & 0.014(3) & 1.499(82) & 10 \\
$f_{11} \approx f_{2}-f_{1}$ 	& 0.50377(13) & 0.061(12) & 1.701  89) & 9 \\
$f_{12} \approx f_{2}+f_{1}$ 	& 2.35493(7)  & 0.032(3) & 1.290  48) & 17 \\
\hline
\hline
\end{tabular}
\end{table}

\begin{figure*}
\centering
\includegraphics [scale=1.4]{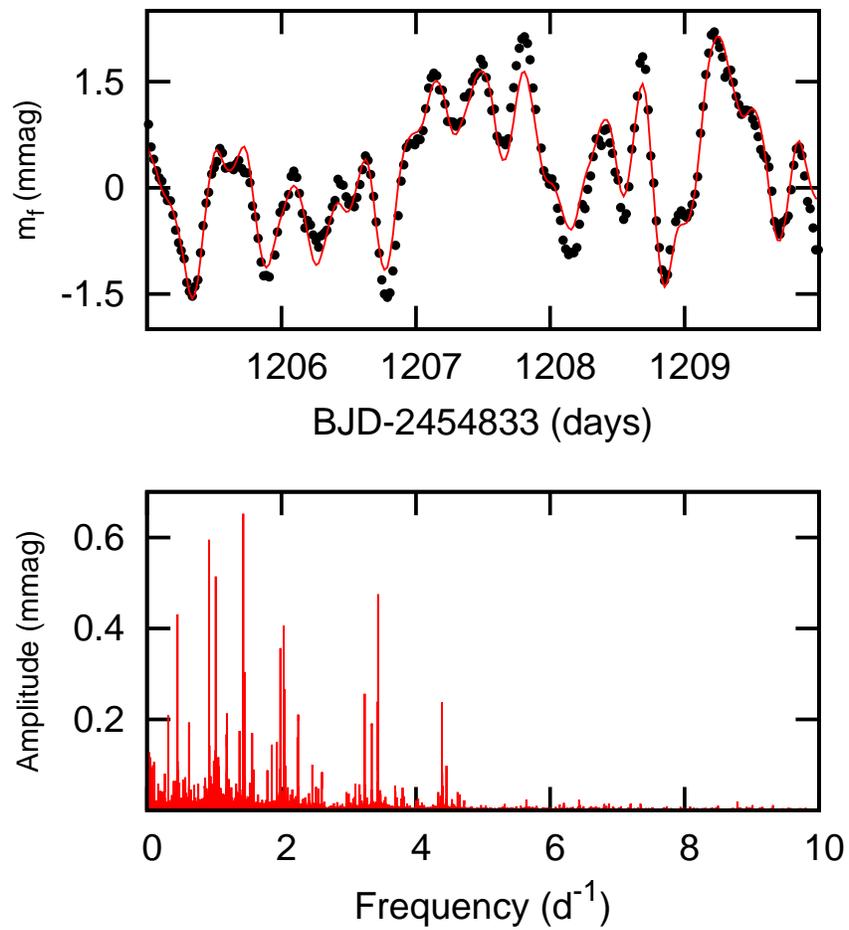} 
\caption{A sample of the {\it Kepler} light curve covering 5d of the original data (top) and
amplitude spectrum (bottom) of KIC\,6462033.}
\label{figm1}
\end{figure*}

\section{The {\it Kepler} Photometry}
 {\it Kepler} data were used to investigate the frequency content of KIC\,6462033 
The {\it Kepler} satellite\footnote{{\it http://kepler.nasa.gov/}} was launched on 2009 March 6 primarily aimed to detect terrestrial and larger planets in the solar neighbourhood by the transit method \citep{bor10, koh10}.  {\it Kepler} has continiously observed the brightnesses of over 100000 stars in a 105 square degree fixed field  for at least 3.5 years. The uninterrupted time series photometry which makes the data ideal for seismic studies has been successful to determine a very large number of  pulsation modes of high accuracy.

KIC\,6462033 was observed with  1- min exposures from BJD~2455156.5156 to BJD~2455182.5060 in SC mode (only Q3.3) and BJD~2454953.5285 to BJD~2455990.9688 in LC modes (Q0-Q16) for a total of 102633 data points. Since most of the observations were obtained in LC mode and SC data frequencies are of limited value  (about 700\,d$^{-1}$) we decided to analyze the LC data collected between the {\it Kepler} commissioning quarters Q0 and Q16. In order to perform frequency analysis, the data were first cotrended the simple aperture Photometry (SAP) fluxes by using cotrending basis vector (CBV) files and  {\tt kepcotrend} task of {\tt PyKE} package \citep{sti12}. After removal of instrumental systematics from the light curve,  final magnitudes were obtained following the formula $m_i =-2.5\log F_i$, where m$_{i}$ refers the magnitude and $F_i$ is the raw SAP flux. We obtained final magnitude (m$_f$) values by subtracting the m$_i$ magnitudes from their fitting polynomial.

{\tt SigSpeC} code \citep{ree07} was used for frequency extraction of  KIC \,6462033. 
The program assigns the spectral significance levels for the discrete Fourier transform (DFT) amplitude spectra 
of time series at randomly time sampling by the fitting formula:

\begin{equation}
\label{e1}
f\left( t \right) = \displaystyle\sum\limits_{i}^n A_{i}\cos \left( 2\pi f t - \phi_{i}  \right)
\end{equation}
where $A_i$ is the amplitude of corresponding frequency, $f$ is the frequency value, and phase angle $\phi_{i}$.

The probability density function (PDF)  of a given DFT amplitude level is determined analytically and false-alarm probability refers to the spectral significance for a certain frequency. The default significance threshold in {\tt SigSpeC} is set at 4.1784 that theoretically corresponds to  $S/N$ = 3.5 \citep{ree07, bre11}. The theoretical Rayleigh resolution is $1/T = 0.000696$~d$^{-1}$.  Although $\gamma$ Dor-type oscillation frequencies are typically detected in the range of 0-5 d$^{-1}$ \citep{bal11} our computations are limited in the range 0 d$^{-1}$-10 d$^{-1}$ to search for frequencies in the region $>$ 5 d$^{-1}$.

The frequency analysis yielded five independent frequencies, several hundreds of combination terms and harmonics up to 2$f_5$. Using SC time series, the first three peaks of high-amplitude were previously reported by \cite{ulu13}.  We also confirm that the light variation is dominated with three frequencies  $f_{1}$=0.92527, $f_{2}$=2.03656 and  $f_{3}$=1.42972 d$^{-1}$. However, the LC data allowed us to detect two other independent terms of much lower amplitude, $f_{4}$=1.02317, and $f_{5}$=3.43850 d$^{-1}$. The lowest frequency that appears significant is $f_{2}$-$f_{1}$=0.50377 d$^{-1}$. On the other hand, the term with the lowest amplitude is detected to be significant in the region around  2 d$^{-1}$. The resulting frequencies of KIC\,6462033  are listed in Table 1, together with their amplitudes, phases and $S/N$ values and uncertainities.

\section{Spectroscopy}
The CCD spectra were used for determination of the atmospheric parameters of KIC\,6462033 and the projection of the rotational velocity $v\sin\,i$. They were obtained with the 2m RCC telescope of the Bulgarian National Astronomical Observatory - Rozhen in three spectral regions: 6510--6610\,\AA\AA, 4810--4910\,\AA\AA\ and 4460--4560\,\AA\AA\ .
The regions were focused on H$\alpha$, H$\beta$, Mg\,II $\lambda$\,4481\,\AA\ lines. The Photometrics AT200 camera with a SITe SI003AB $1024 \times 1024$ CCD chip,
($24\,\mu {\rm m}$ pixels) was used in the third camera of the Coud\'e spectrograph to provide spectra with a typical resolution R = 32\,000 and S/N ratio of about 40. 
The instrumental profile was checked by using the comparison spectrum so that its FWHM was about 0.2~\AA. {\tt IRAF} standard procedures were used for bias subtracting, 
flat-fielding and wavelength calibration. The final spectra were corrected to the heliocentric wavelengths.

Model atmospheres were calculated under {\tt ATLAS\,12} code.
The {\tt VALD} atomic line database (\citealt{kprsw99}), which also contains
\citet{Kurucz93} data, was used to create a line list for the synthetic spectra. 

The synthetic spectra were obtained by using the code {\tt SYNSPEC}  (\citealt*{hlj94}, \citealt{Krticka98}). 
We accepted the microturbulence to be 2\,km\,s$^{-1}$. 
The computed spectra were convolved with the instrumental profile by a Gaussian of 0.2\,\AA ~FWHM 
and rotationally broadened to fit the observed spectra.

\begin{figure*}
\centering
\includegraphics[scale=1.5]{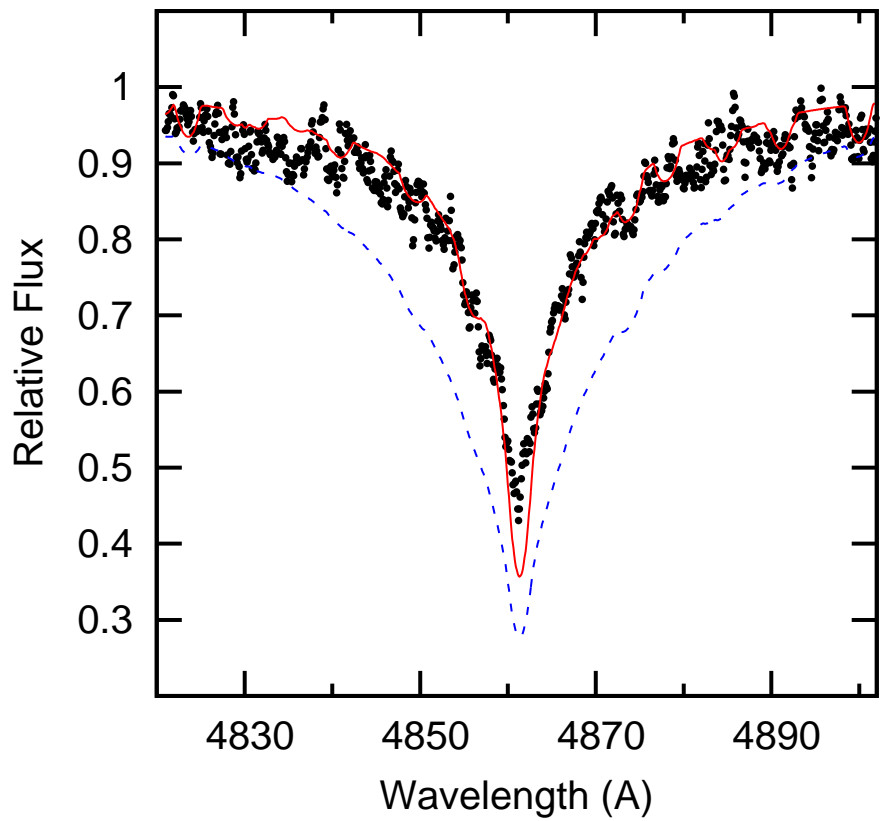} 
\caption{H$\beta$ line (dots) fitted with a model with $T_{\rm eff}=7150~{\rm K}$, $\log g=4.3$ 
(solid line). The model proposed by KIC - $T_{\rm eff}=8390~{\rm K}$, $\log g=4.3$ is given by dashed line.}
\label{fsyn}  
\end{figure*}

The best fit of H$\beta$ and H$\alpha$ was obtained for the atmosphere model with $T_{\rm eff}=7150~{\rm K}$, 
$\log g=4.3$. We used Mg\,II $\lambda$\,4481\,\AA\, line for the determination of the projected rotational velocity. 
The match between the synthetic and observed profile gave as a result $v\sin\,i=90~{\rm km\,s^{-1}}$. In Fig. \ref{fsyn} 
it can be seen the best fit of H$\beta$ and the fit with the model given by {\tt Kepler} Input Catalogue (hereafter KIC) \citep{uyt11}. The values given by KIC - $T_{\rm eff}=8390~{\rm K}$, $\log g=4.3$ differ from our results concerning the temperature but it is not surprisingly.
As \cite{tkac12} mentioned in their spectroscopic work on $\gamma$ Doradus stars the spectroscopically derived temperatures 
in some cases disagreed with the KIC values.  

\section{Conclusions}
We have presented both frequency and first spectrum analysis of the $\gamma$ Doradus star KIC\,6462033 observed by the  {\it Kepler} satellite.  In order to extract frequencies from the LC {\it Kepler} data, the frequency analysis was performed using the software package {\tt SigSpeC} \citep{ree07}.  Our results are in a good agreement with previous analyses of the SC {\it Kepler} data \citep{ulu13}. We confirm that the light curve of  KIC\,6462033 is dominated by three modes with frequencies  $f_{1}$= 0.92527, $f_{2}$= 2.03656, $f_{3}$= 1.42972 d$^{-1}$. Besides these three modes, two additional frequencies which have much lower amplitudes are detected: $f_{4}$= 1.02317, and $f_{5}$= 3.43850 d$^{-1}$ together with their harmonics and combination terms. No significant peaks are appeared in the region $>$ 5 d$^{-1}$. As a result,  KIC\,6462033 pulsates in the frequency range ($\leq$5 d$^{-1}$) of  $\gamma$ Dor type variables \citep{gri10}.

Since ground-based spectroscopy is needed to complete the obtained results by using  {\it Kepler}  photometry, we also carried out high dispersion spectroscopic observations to derive the fundamental parameters and projected rotational velocity for KIC\,6462033. By fitting the Balmer line profiles and Mg\,II $\lambda$\,4481\,\AA\, line between the synthetic and  observed spectra we obtained $T_{\rm eff}=7150~{\rm K}$, $\log g=4.3$, $v_{mic}$= 2\,km\,s$^{-1}$ and $v\sin\,i=90~{\rm km\,s^{-1}}$ respectively. We find that the values of effective temparature and surface gravity listed in the KIC are not overlapped with the ones derived from our ground-based spectra.

In  $\gamma$ Dor stars,  spherical harmonic degrees ($l$)  an idendification of the radial order ($n$) of observed frequencies can be determined with frequency ratio method (FRM) \citep{moya05, su05}. This method was particularly developed for $\gamma$ Dor variables that show at least three $g$-mode pulsation frequencies in the asympthotic regime. The method also provides an estimate of the integral of the Brunt-V\"{a}is\"{a}l\"{a} frequency weighted over the stellar radius along the radiative zone if we assume that all exicited modes in the star correspond to the same spherical harmonic degree $l$ with $m$=0. However, it should be noted that the FRM is only applicable and reliable for objects with rotational velocities up to  70~km\,s$^{-1}$ due to the effect of rotation on the observational Brunt-V\"{a}is\"{a}l\"{a} frequency \citep{su05}. Considering that the rotational velocity of KIC\,6462033 is over the limit of validity of the FRM we therefore didn't attempt mode identification for KIC\,6462033 with FRM. In conclusion, we expect that more high-resolution spectroscopic observations may enable us to identify at least some of modes and our present work will support in future asteroseismic studies of this star.

\section*{Acknowledgments}
The authors acknowledge the whole Kepler team for providing the
unprecedented data sets that make these results possible. This paper
includes data collected by the Kepler mission. Funding for the
Kepler mission is provided by the NASA Science Mission directorate.
CU wishes to thank the South African National Research Foundation (NRF) for
the award of NRF Multi-Wavelength Astronomy Research Programme (MWGR), 
Grant No: 86563 to Reference: MWA1203150687. IS and II gratefully acknowledge the partial support
from Bulgarian NSF under grant DO 02-85. BU is supported by the project numbered
HDYF-078. The authors also sincerely thank Garith Dugmore for ICT support during the analysis. This work made use of PyKE (Still \& Barclay 2012), a software package for the reduction and analysis of Kepler data. This open source software project is developed and distributed by the NASA Kepler Guest Observer Office.This study made use of IRAF Data Reduction and Analysis System and the Vienna Atomic Line Data Base (VALD) services.

\end{document}